# High-Level Requirements Management and Complexity Costs in Automotive Development Projects: A Problem Statement

Tim Gülke[1], Bernhard Rumpe[1], Martin Jansen[2], and Joachim Axmann[2]

[1] Software Engineering, RWTH Aachen University
[2] Volkswagen AG, Wolfsburg

**Abstract.** Effective requirements management plays an important role when it comes to the support of product development teams in the automotive industry. A precise positioning of new cars in the market is based on features and characteristics described as requirements as well as on costs and profits. [Question/problem] However, introducing or changing requirements does not only impact the product and its parts, but may lead to overhead costs in the OEM due to increased complexity. The raised overhead costs may well exceed expected gains or costs from the changed requirements. [Principal ideas/results] By connecting requirements with direct and overhead costs, decision making based on requirements could become more valuable. [Contribution] This problem statement results from a detailed examination of the effects of requirements management practices on process complexity and vice versa as well as on how today's requirements management tools assist in this respect. We present findings from a joined research project of RWTH Aachen University and Volkswagen.

## 1 Today's Requirements Management in Automotive Practice

The automotive industry is facing several challenges ranging from entirely new engine concepts to customer-configurable infotainment systems and networks of computers and infrastructure. The trend of increasing product complexity has not yet been stopped and is still gaining speed [9], which also leads to growing complex structures within the companies [12]. For the automotive industry, Schleich et al. [16] already linked increasing numbers of variants with rising complexity and overhead costs.

Requirements management plays a vital role by providing supportive processes and tools for the employees engaged in development activities [8,13]. Particularly in the process of defining a product's characteristics – e.g., what infotainment features will be available to the customer, how many different types of engines for which sort of fuels, or how many passengers the car will be designed

for – and later changes to those, requirements management aids in the engineers day to day work. For the last few years, ideas which have been developed in theory have proven themselves functional in practice, although much work still remains to be done [18]. This refers, e.g., to the application of templates in requirements elicitation, the usage of clear and non-ambiguous words, traceability in general and the inclusion of suppliers into the requirements work [6,15]. When two or three decades ago a single group of employees was able to keep track of the requirements for a car with pen and paper and in their heads, nowadays collected information is spread through countless documents, systems, and people. The evolutionary step from vehicle platforms to modules and modular toolkits makes it even more difficult, since now links between requirements and parts are not limited to one vehicle anymore. Requirements management can therefore be seen as a measure to handle the increasing complexity by providing a way of keeping all necessary information connected. It enables engineers to estimate the impact of proposed changes and equips the project leaders with powerful tools to track status.

The underlying concept of requirements management is traceability, which means the connection of different artifacts throughout one or multiple projects. Therefore, a requirements management tool can only be as good as the level of traceability it operates on when it comes to impact analysis of changes or additional requirements. So far, traceability connects most product-related things like parts, functions, all kind of documents and specifications, scenarios and tests with requirements. The amount to which this is done differs in companies and also in projects. The complexity of electronic systems in vehicles forces the automotive industry to maintain a high level of traceability within their projects [7]. This is why it is current practice to be able to estimate the costs of changes on a very detailed level, knowing the impact of a proposed change by tracing all connected artifacts.

Today, there are several programs available on the market supporting development teams in eliciting, organizing, tracing, linking, and generally managing requirements. Tools like IBM DOORS, Borland CaliberRM, Jama Contour, and others provide the ability to describe requirements in a specific way, implement hierarchies and most possess modeling-functionality for the underlying structure [5]. However, they're all limited to product-centric models and do not provide any way of including costs originating in processes far from the product (or even costs at all).

## 2   Requirements and Costs

Changes in requirements or the introduction of new requirements (regardless in what stage the current project is in) lead to three different types of costs:

1. **Investment Costs:** Costs originate from necessary investments into the development of a product and its parts. This includes, e.g., the purchase of tools and machines as well as the production of prototypes.

2. **Direct Costs:** These are the later internal "pricetags" on parts or whole systems, whether they're bought from a supplier or made in-house when it comes to the production of the car. There are usually targets defined for every individual part or function to stay within a defined price-range for the whole car with which it is placed on the market.
3. **Overhead/Indirect Costs:** Overhead costs are costs which occur in the production-phase of the car that cannot be related to a definite cost object (i.e. a vehicle sold on the market). They're generated by employees filling out excel-sheets, making phone-calls, etc.

If a requirement is added or changed, two things happen: First, additional investment costs are generated because a new or different feature is included into the car. Reasons behind the requirement can be manifold and range from competitors providing a new function with their vehicles that has to be matched to regulatory/legal problems. Second, overhead costs may rise due to an increased complexity in the processes of the company [16]. The estimation of investment costs for a proposed change is done very accurately, but mostly relies on the knowledge of the engineers regarding the type of the change. This slows down the decision process which then again slows down the early phases of a vehicle development project. Decision-makers are left with three choices when it comes to predicting the overall costs caused by a requirement:

1. Huge manual effort can be put into figuring out which departments are affected by a change (purchasing department? engineering? marketing? which ones exactly?) and then ask each of those to estimate the amount of work needed. These two steps are time- and cost-intensive and the results are not guaranteed to be exact.
2. Another way is to use a fixed amount of money based on prior experiences with similar changes. This might cause problems, since it's unclear whether this amount is accurate to the actual costs, but it's quick and feasible.
3. Last, those costs can simply be added to the affected departments overhead costs and not be counted against a project's budget.

While investment costs can at least be estimated, the prediction of the change of complexity in OEMs (and suppliers) is difficult and rarely done. A new variant, caused by a changed requirement, leaves only small traces in the company – e.g., one more line to be added to an MS Excel sheet, one more item to be synchronized between two systems, one more line in a report, etc. – and mostly causes administrative work [4]. It is estimated though, that if the number of variants are doubled, overhead costs rise 20%-30% because of increased complexity [19]. Strikingly it is the combined number of small steps that can cause this increase, but they are not part of the decision process, since it is difficult to predict where exactly what amount of additional work is caused [16].

Complexity's impact on products is currently under research and approaches are being proposed [11], some work is done with regard to complexity [12] and of course many new developments in the field of requirements management are being published [10], although many focus on software-only projects [14]. It seems promising to combine these different areas of research for practical use and extend the current focus of the product in requirements management to processes and their complexity. Almefelt et al. [2] for example already recommend the conduction of a cost/benefit-analysis for requirements changes.

## 3 Example

Automotive OEMs follow a combined sequential/iterative process-model during the development of a new vehicle. This leads to an early declaration of requirements in a so-called *product definition phase*, where different business units collect, exchange and adjust their requirements for the new car. Based on an early bill of materials, costs are estimated for the realization of the requirements. These costs include necessary investment costs and expected direct costs of the car in later production. Requirements may lead to a decrease of direct costs, e.g., by making a single part of the car available in two or more different variants, some applying inexpensive materials, the others with the standard ones and using these accordingly in different variants of the car (e.g., in different "lines" or brands). If the installation rate of the lower-cost part is high enough, revenues will be raised.

However, the new variant of the part has not only to be constructed or programmed, bought from suppliers, stored in factories, databases, etc. but to be maintained in different systems and processes as an artifact – and these make up of most for the overhead costs. It has already been published in Schleich et al. [16] that with an increased variability, overhead costs rise in the field of production and logistics, but the figures of how this rising variability combined with construction kits and platforms affects costs in product development and change processes cannot yet be answered. It is therefore to be suspected that changed or new requirements might partially lead to costs that exceed revenues gained from them. If an accepted methodology and software were available that estimated how a requirement affects the companies complexity, the raise of overhead costs could at least be controlled. It can be assumed that certain topics might be decided differently, if complexity costs were considered in the decision-making process. Last, the approach would allow a cleanup of variants with complexity costs that are significantly higher than their revenues and thus lower a company's overhead costs.

## 4 Extending Traceability

Today's automotive companies are confronted with increasing complexity not only in it's products, but also in their internal organizations. This is seldom considered when it comes to requirements changes during vehicle development projects. Doing this manually for each change is error-prone and cost-intensive. Requirements management tools should widen their focus from a straight product view towards a process view that includes all aspect of a company since very few tools provide the ability to model processes at all or they do only focus on automating simple tasks and routines.

The key to this problem might be the thorough modeling of corporate structures and artifacts. Making knowledge of this kind available to software will enable it to consider far more aspects of decisions than it does now. But so far, the creation of models from a company's artifacts decoupled from a concrete software project is seldom done, since the benefit is not immediately visible. Even the formal description of processes will only be done if the need arises to automate some parts of the process or in optimization projects.

But first of all, requirements management tools need to implement cost-structures (e.g., from product data management systems) and connect them with their data models. This will enable decision-makers to anticipate how a certain change would affect direct costs. Afterwards, process-engineering tools like ARIS [1] can be connected to extend the decision-process by the inclusion of the affected processes. Once all this data is present, modeling of artifacts inside the processes can begin, providing an even deeper insight into how, e.g., a new variant will be processed throughout the whole company.

The mentioned topics can be seen as an extension to the already powerful concept of traceability. Making not only parts or documents, but all artifacts of a company traceable, will enable decision-makers to estimate investment costs faster and predict the change of complexity. This can only be done if tools are available that have the ability to include these artifacts or at least be able to communicate with systems that do.

A company-wide repository for models of artifacts like processes or documents would need a standardized description language, which is able to both capture the models and set them into context with each other. Efficient modeling tools need to be available as well that support model developers in creating those models fast enough to keep up with the pace of change in a company. Next, requirements management tools would need to use the available models and their contexts and wave them into their own traceability model – and maybe even provide a way other tools could reuse those models.

The research area of semantic networks already provides languages and concepts to capture information as described above. Connecting these with the powerful tools available in the requirements management world might prove valuable. Languages like OWL/RDFS which are thoroughly documented [3] could be used to construct a knowledge repository that requirements management tools could use. Approaches providing a way of automated ontology creation for the gathering of this semantic data might be helpful [17].

Knowing the financial benefit beforehand is difficult, since the costs that are going to be addressed are not traceable so far – otherwise this problem would not exist. Therefore, only the careful introduction of an approach like this will definitely show its benefits. But since the automotive world is getting more complex every day with a widened portfolio in brands and products and more detailed markets being all deeply connected, it needs the ideas and concepts traceability and requirements management provide.

There is no denying that more research is needed on how requirements and costs play together. Also, a solution for the efficient and easy modeling of process artifacts is necessary, as well as how to use that knowledge in a requirements management tool. To come to an end, not only might this problem be an automotive-industry specific one, but it could also be extended into other domains.